\documentclass[12pt]{iopart}

\usepackage{amsthm}
\usepackage{amsfonts}
\usepackage{graphicx}

\theoremstyle{plain}
\newtheorem{theorem}{Theorem}
\newtheorem{prop}{Proposition}
\newtheorem{conj}{Conjecture}
\newtheorem{lem}{Lemma}

\theoremstyle{definition}
\newtheorem{defn}{Definition}

\newcommand{\U}{{\bf U}}

\newcommand{\M}{{\bf M}}
\newcommand{\B}{{\bf B}}
\newcommand{\A}{{\bf A}}
\newcommand{\vv}{{\bf v}}
\newcommand{\uu}{{\bf u}}
\newcommand{\ee}{{\bf e}}
\newcommand{\bN}{{\mathbb N}}

\begin{document}

\title[Diagonal approximation of the CUE form factor]
{Diagonal approximation of the form factor of the unitary group}
\author{G Berkolaiko}
\address{Department of Mathematics,
Texas A\&M University, College Station TX 77840-3368, USA}
\ead{Gregory.Berkolaiko@math.tamu.edu}
\date{April 2, 2001}

\begin{abstract}
  The form factor of the unitary group $U(N)$ endowed with the Haar
  measure characterizes the correlations within the spectrum of a
  typical unitary matrix.  It can be decomposed into a sum over pairs
  of ``periodic orbits'', where by periodic orbit we understand any
  sequence of matrix indices.  From here the diagonal approximation
  can be defined in the usual fashion as a sum only over pairs of
  identical orbits.  We prove that as we take the dimension $N$ to
  infinity, the diagonal approximation becomes ``exact'', that is
  converges to the full form factor.
\end{abstract}


\pacs{0545M, 0210D, 0250G, 0510G, 0270H}

\maketitle


\section{Introduction}

One of the central questions in the quantum theory of classically
chaotic systems is whether the fluctuations in the spectra follow the
predictions of random matrix theory (RMT).  The spectral form factor, which
is the Fourier transform of the two-point correlation function, is
often used to characterize these fluctuations.

The advantage of the form factor is that it has a convenient expansion
in terms of pairs of periodic orbits of the classical system.  This
expansion is an application of trace formulae (see \cite{CPS98} for a
review) and has been a starting point for many investigations.  The
progress of understanding the role of periodic orbits in the universal
behavior of the form factor of different systems is marked by such
milestones as the diagonal approximation \cite{Ber85}, the first
off-diagonal contribution \cite{Sie02,SR01} and the most recent preprint
\cite{MHBHA05} outlining the complete expansion.  Quantum graphs were
also used \cite{BSW03} as a mathematical toy model with a
simple-to-understand periodic orbit structure and an exact trace
formula.

In a recent paper, Degli Esposti and Knauf \cite{DK04} proposed to use
the ultimate mathematical toy model for the random matrix conjecture
--- the random matrix theory itself.  The authors argued that if one
can perform a step-by-step expansion of the RMT form factor in terms
of families of periodic orbits, it would greatly facilitate analogous
expansions for other systems.

Degli Esposti and Knauf \cite{DK04} studied the diagonal approximation
to the form factor of one the random matrix ensembles, the Circular
Unitary Ensemble (CUE).  Their results, however, depend on a
combinatorial conjecture, which is still unproven.  The purpose of
this letter is to present an alternative proof for their main result;
our proof is unconditional and is significantly simpler.  Our proof is
based on the observation that the diagonal approximation can be
expressed as a trace of certain doubly stochastic matrix.  The
asymptotic properties of the diagonal approximation can then be
related to the spectral gap of the doubly stochastic matrix, which was
previously studied by the author in \cite{B01}.  

The structure of the letter is as follows.  In Section~\ref{sec:CUE}
we introduce the Circular Unitary Ensemble and the form factor.  We
follow with reviewing the main result of \cite{DK04} and stating our
main result in Section~\ref{sec:results}.  The relevant properties of
doubly stochastic matrices are reviewed in Section~\ref{sec:proof}.
Here we also prove our main result which follows quite simply from a
technical lemma on the asymptotics of the second largest eigenvalue.
This lemma is proved in Section~\ref{sec:lemma_proof} and we conclude
with comparing our results with those of \cite{DK04} in
Section~\ref{sec:discussion}.

\section{Circular Unitary Ensemble and its form factor}
\label{sec:CUE}

The Circular Unitary Ensemble is the unitary group U($N$) equipped
with the Haar measure:
\begin{defn}
  CUE$(N)$ is defined as the ensemble of all unitary $N\times N$
  matrices endowed with the probability measure that is invariant
  under every automorphism
  \begin{equation}
    \label{eq:autom}
    \U \mapsto {\bf VUW },
  \end{equation}
  where ${\bf V}$ and ${\bf W}$ are any two $N\times N$ unitary
  matrices.
\end{defn}

The form factor is a function describing the statistical properties of
the spectrum of $\U$.  It is defined as the Fourier transform of the
two-point spectral correlation function, $R_2(x)$,
\begin{equation}
  \label{eq:R2}
  R_{2,N}(x) 
  = \left\langle\frac{1}{N} 
    \sum_{j,k=1}^N\delta(x + (\phi_j-\phi_k)N)\right\rangle_N,
\end{equation}
where $\exp(2\pi i \phi_j)$ are the eigenvalues of $\U$ and the
averaging is performed over the CUE$(N)$.
The form-factor of the unitary ensemble is thus given by
\begin{equation}
  \label{eq:FF}
  K_N(t) = \frac{1}{N} \left\langle|\Tr(\U^t)|^2 \right\rangle_N 
  \qquad (t\in{\mathbb Z}),
\end{equation}
The well-known formula for $K_N(t)$ is
\begin{equation}
  \label{eq:FF_ans}
  K_N(t) = \left\{
    \begin{array}{lcl}
      N,&t=0\\
      |t|/N,&0<|t|\leq N\\
      1,&N<|t| 
    \end{array}
    \right. \qquad (t\in{\mathbb Z}).
\end{equation}
This formula can, in particular, be derived by expanding the trace in
(\ref{eq:FF}),
\begin{equation}
  \label{eq:FF_trace_expanded}
    K_N = \frac{1}{N}\sum_{i_1,\ldots,i_t=1}^N \sum_{j_1,\ldots,j_t=1}^N
    \left\langle\prod_{k=1}^t U_{i_ki_{k+1}}U_{j_kj_{k+1}} \right\rangle_N,
\end{equation}
where the convention $i_{t+1}=i_1$ is assumed here and throughout the
manuscript.  The above expression is analogous to the trace-formula
expansions of the form factor in other systems, with ${\bf
  i}=(i_1,\ldots,i_t)$ and ${\bf j}=(j_1,\ldots,j_t)$ playing the role
of the periodic orbits.  It can be shown (see equation (\ref{eq:avU})
below) that only the terms with ${\bf j}$ being a permutation of ${\bf
  i}$ contribute to the form factor, i.e.
\begin{equation}
  \label{eq:FF_trace_form}
  K_N = \frac{1}{N}\sum_{i_1,\ldots,i_t=1}^N \sum_{\sigma\in S_t}
  \left\langle\prod_{k=1}^t U_{i_ki_{k+1}}U_{i_{\sigma(k)}j_{\sigma(k+1)}} 
  \right\rangle_N,
\end{equation}
where $S_t$ is the group of all $t$-permutations (the symmetric
group).  This step is analogous to only considering pairs of orbits with a
small action difference.  The approximation to
(\ref{eq:FF_trace_form}) that restricts the double sum even further,
to pairs of periodic orbits that are equal (up to a cyclic shift), is called
the {\em diagonal approximation}
\begin{equation}
  \label{eq:FF_diag}
  K_{1,N} = \frac{t}{N}\sum_{i_1,\ldots,i_t=1}^N 
  \left\langle\prod_{k=1}^t|U_{i_ki_{k+1}}|^2 \right\rangle_N,
\end{equation}
where the factor $t$ comes from the number of possible shifts of an
orbit.  In this letter we compare the form factor $K_N(t)$ with its
diagonal approximation.

The present article was inspired by a study by Degli Esposti and
Knauf, \cite{DK04}.  They proved that, assuming validity of a certain
combinatorial conjecture, the diagonal approximation $K_{1,N}$
converges to $t/N$ uniformly in $t/N$ as $N\to\infty$.  The
convergence is proved in a closed interval which is bounded away from
zero.  The combinatorial conjecture requires a more thorough
introduction and will be delayed to a later section.

Here we propose an alternative approach which exploits the fact that
the sum in $K_{1,N}$ is the trace of a doubly stochastic matrix.  Our
proof is unconditional, i.e. it does not rely on the conjecture, and
the interval of convergence is extended.

\section{Averages over unitary group and the main result}
\label{sec:results}

The averages of products of the elements of matrices $\U$ have been
studied by various authors, \cite{Sam80,Ull64,Mel90,BB96}.  Their main
result is
\begin{equation}
  \label{eq:avU}
  \langle U_{a_1b_1}\ldots U_{a_sb_s} 
  U^*_{\alpha_1\beta_1}\ldots U^*_{\alpha_t\beta_t} \rangle_N = 
  \delta_t^s \sum_{\sigma,\pi\in S_t} V_N(\sigma^{-1}\pi)
  \prod_{k=1}^{t}
  \delta_{a_k}^{\alpha_{\sigma(k)}} \delta_{b_k}^{\beta_{\pi(k)}},
\end{equation}
where $S_t$ is the symmetric group of permutations of the set
$\{1,\ldots, t\}$, $\delta_k^n$ is the Kronecker delta and the
coefficient $V_N(\sigma^{-1}\pi)$ depends only on the lengths of
cycles in the cycle expansion of $\sigma^{-1}\pi$.

To introduce the results of \cite{DK04} we need the following definitions.
\begin{defn}
  Let $\sigma\in S_t$ and $i\in\{1,2,\ldots,t\}$.  Then a {\em cycle} of
  $i$ with respect to $\sigma$ is the (finite) set $\{\sigma^k(i)
  \colon k\in\bN\}$.  The {\em cycle length} of $i$ is the number of
  elements in the cycle of $i$.  The {\em rank} $|\sigma|$ of a
  permutation $\sigma$ is the number of disjoint cycles induced by
  $\sigma$.  By $|\sigma \vee \rho|$ we denote the number of
  disjoint sets of the form $\{\rho^n\sigma^k(i) \colon k,n\in\bN\}$.
  By $\tau$ we will denote the circular permutation $(1,2,\ldots,t)$.
  If $\sigma\in S_t$, $\sigma_+$ will denote $\tau^{-1}\sigma\tau$.
\end{defn}

\begin{conj}[Degli Esposti -- Knauf \cite{DK04}]
  \label{conj}
  There exists a constant $C_1$ such that for all $t\leq N \in \bN$
  \begin{equation}
    \label{eq:conj}
    \sum_{\pi\in S_t} V_{N}\left(\pi^{-1}\sigma\right)
    N^{|\pi\vee\sigma_+|} 
    \leq C_1 N^{|\sigma\vee\sigma_+|-t}  \qquad \sigma\in S_t,
  \end{equation}
\end{conj}

\begin{theorem}[Degli Esposti -- Knauf \cite{DK04}]
  \label{thm:MDE_AK}
  Assuming Conjecture~\ref{conj} is true, the form factor $K_N$ is
  approximated by the diagonal contribution in the following sense:\\
  For all $\epsilon>0$
  \begin{equation}
    \label{eq:convergence1}
    |K_N(t) - K_{1,N}(t)|\to0 \qquad \mbox{ as } N\to\infty
  \end{equation}
  uniformly in $\tau=t/N\in[\epsilon, (1-\epsilon)e/C_1]$.
\end{theorem}

Our main result is a version of the above theorem that drops the
dependence on the Conjecture and extends the interval of convergence.
\begin{theorem}
  \label{thm:main_result}
  For all $\epsilon>0$
  \begin{equation}
    \label{eq:convergence2}
    |t/N - K_{1,N}(t)|\to0 \qquad \mbox{ as } N\to\infty
  \end{equation}
  uniformly in $\tau=t/N\in[\epsilon, \epsilon^{-1}]$.  Thus, uniformly in
  $\tau=t/N\in[\epsilon, 1]$, 
  \begin{equation}
    \label{eq:convergence3}
    |K_N(t) - K_{1,N}(t)|\to0 \qquad \mbox{ as } N\to\infty.
  \end{equation}
\end{theorem}

\section{Doubly stochastic matrices and the proof of Theorem
  \ref{thm:main_result}}
\label{sec:proof}

\begin{defn}
  An entry-wise nonnegative $N\times N$ matrix $\M$ is called {\sl
    doubly stochastic} if
  \begin{equation}
    \sum_{i=1}^N M_{i,j} = 1\quad  \forall j
    \qquad\mbox{ and }\qquad 
    \sum_{j=1}^N M_{i,j} = 1\quad  \forall i.
  \end{equation}  
  The set of all such $N\times N$ matrices we denote by DS$(N)$.
  DS$(N)$ is a monoid with respect to the matrix multiplication, in
  particular if $\A$ and $\B$ are doubly stochastic
  then so is $\A\B$.
\end{defn}

It follows directly from the definition that the vector
$\ee=(1/N,\ldots,1/N)$ is both a left and a right eigenvector of 
any matrix from DS$(N)$.  The corresponding eigenvalue 1 is the
largest (by modulus) one.  The following observation shows the
importance of the asymptotics of the second largest eigenvalue.

If $\U$ is unitary then the matrix $\M$ defined by
\begin{equation}
  \label{eq:trans_prob}
  M_{i,j} = |U_{i,j}|^2.
\end{equation}
is doubly stochastic.  It is then easy to see that
\begin{equation}
  \label{eq:trM}
  \Tr \M^t = \sum_{j=1}^N (\M^t)_{j,j} 
  = \sum_{i_1,\ldots,i_t=1}^N \prod_{k=1}^t \left|U_{i_ki_{k+1}}\right|^2,
\end{equation}
and, therefore, the diagonal approximation $K_{1,N}$ is equal to
\begin{equation}
  \label{eq:diag_trM}
  K_{1,N} = \frac{t}{N} \langle \Tr \M^t \rangle_N,
\end{equation}
where the averaging is performed with respect to the measure induced
on doubly stochastic matrices by the correspondence
(\ref{eq:trans_prob}).  In \cite{B01} it was speculated that the
second largest eigenvalue of such random matrices\footnote{an
  extensive discussion of properties of such random matrices is
  contained in \cite{Zyc03}} $\M$ decay, on average, as $N^{-1/2}$ .
If this is indeed so, we immediately get
\begin{equation}
  \label{eq:intuition}
  \langle \Tr \M^t \rangle_N = 1 + \sum_{i=2}^N \langle \lambda_i^t \rangle 
  \approx 1 + (N-1) N^{-t/2} \to 1,
\end{equation}
for $t>2$, which would prove Theorem~\ref{thm:main_result}.  While
the conjecture of \cite{B01} is still unproven, we can use the
technique of \cite{B01} to achieve the same result without directly
calculating the eigenvalues of $\M$.

Since the eigenvalues of $\M$ lie in the unit circle, it is
intuitively clear that $\Tr \M^t = \sum \lambda_i^t$ should be a
decreasing function in $t$, at least ``on average''.  It is therefore
tempting to show that $\langle \Tr \M^t \rangle_N \to 1$ for some {\em
  fixed} $t$ as $N\to\infty$ and then conclude that this property
still holds if we send $t$ to infinity as well.  Indeed, this would
work if the eigenvalues of $\M$ were positive real numbers, which is
generically not the case for matrices $\M$.  To circumvent this
difficulty, we form a matrix
\begin{equation}
  \label{eq:definitionA}
  \A = \M^T\M,
\end{equation} 
which is symmetric, doubly stochastic and positive definite.  Now we
can estimate the eigenvalues of $\M$ through the eigenvalues of $\A$.
\begin{prop}[\cite{B01}]
  Second largest eigenvalues of $\M$ and $\A$ satisfy
  \begin{equation}
    \label{eq:ineq}
    |\lambda_2(\M)|^2 \leq \lambda_2(\A).
  \end{equation}
\end{prop}

\begin{proof}
  Let $\vv$ be the eigenvector of $\M$ corresponding to $\lambda_2(\M)$
  and orthogonal to $\ee$.  Such eigenvector always exists: if
  $\lambda_2(\M)\neq 1$, then we write
  \begin{equation}
    \label{eq:orth}
    (\vv, \ee) = (\vv, \M^{T}\ee) = (\M\vv, \ee) = \lambda (\vv, \ee),
  \end{equation}
  and therefore $(\vv, \ee)=0$.  If $\lambda_2(\M)=1$, the geometric
  multiplicity of $1$ is equal to its algebraic multiplicity (see,
  e.g.~\cite{Gan59}) and we can find $\vv$ by orthogonalization.
  
  Then, since all eigenvectors of $A$ are orthogonal, we have
  \begin{equation}
    \label{eq:lem_proof}
    \lambda_2(\A) = \max_{|\uu|=1,\ (\uu,\ee)=0} (\A \uu, \uu) 
    \geq (\A \vv, \vv) = (\M \vv, \M \vv) = |\lambda_2(\M)|^2,
  \end{equation}
\end{proof}

Now the idea is to find the power $k$ such that
$\langle\lambda(\A)^k\rangle_N$ decay sufficiently fast.  It turns out
that $k=3$ is enough.
\begin{lem}
  \label{lem:eigA}
  The Haar probability measure on the unitary group induces a measure
  on symmetric doubly stochastic matrices $\A$ via (\ref{eq:trans_prob})
  and (\ref{eq:definitionA}).  With respect to this measure
  $\langle\lambda_2(\A)^3\rangle_N = O(N^{-2})$.
\end{lem}
The proof is rather technical and will be given in a later section.
This lemma enables us to give a simple proof of our main result.

\begin{proof}[Proof of Theorem~\ref{thm:main_result}]
  We show that $\langle\Tr\M^t\rangle \to 1$.  If $\tau\geq\epsilon$
  and $N$ is sufficiently large then $t = \tau N \geq 6$, leading to
  \begin{eqnarray*}
    \label{eq:proof_main}
    \fl    |\langle\Tr\M^t\rangle - 1| 
    = \left|\left\langle \sum_{k=2}^N \lambda_k(\M)^t\right\rangle\right|
    \leq \sum_{k=2}^N \left|\left\langle\lambda_k(\M)^t\right\rangle\right| 
    \leq \sum_{k=2}^N \left\langle |\lambda_k(\M)|^t \right\rangle\\
    \leq (N-1) \left\langle |\lambda_2(\M)|^t \right\rangle
    \leq (N-1) \left\langle |\lambda_2(\M)|^6 \right\rangle\\
    \leq (N-1) \left\langle |\lambda_2(\A)|^3 \right\rangle = O(N^{-1}).  
  \end{eqnarray*}
  Now we use equation~(\ref{eq:diag_trM}) to obtain the sought after
  conclusion.
\end{proof}

\section{Proof of Lemma~\ref{lem:eigA}}
\label{sec:lemma_proof}

We start with the trace of $\A^3$,
\begin{equation}
  \label{eq:trA}
  \Tr\A^3 = 1 + \sum_{k=2}^N \lambda_k(\A)^3 \geq 1 + \lambda_2(\A)^3, 
\end{equation}
since all eigenvalues of $\A$ are non-negative real numbers.  Thus we
can estimate $0 \leq \lambda_2(\A)^3 \leq \Tr\A^3 - 1$.  Writing out
the trace we get
\begin{eqnarray*}
  \label{eq:trA_expand}
  \fl\Tr\A^3 = \Tr(\M^T\M\M^T\M\M^T\M) \\
  = \sum_{i,j,k,l,m,n=1}^N
  (\M^T)_{i,j}(\M)_{j,k}(\M^T)_{k,l}(\M)_{l,m}(\M^T)_{m,n}(\M)_{n,i} \\
  = \sum_{i,j,k,l,m,n=1}^N
  |U_{j,i}U_{j,k}U_{l,k}U_{l,m}U_{n,m}U_{n,i}|^2.
\end{eqnarray*}
Now we average over the unitary group,
\begin{equation}
  \label{eq:estim_average}
  \langle\lambda(\A)^3\rangle_N + 1
  \leq \sum_{i,j,k,l,m,n=1}^N \left \langle
    |U_{j,i}U_{j,k}U_{l,k}U_{l,m}U_{n,m}U_{n,i}|^2 
  \right\rangle_N.
\end{equation}
At this point we can apply equation (\ref{eq:avU}) with
$\alpha_k=a_k$ and $\beta_k=b_k$ with $1\leq k \leq6$.  The indices
are ${\bf a}=(j,j,l,l,n,n)$ and ${\bf b}=(i,k,k,m,m,i)$.  We denote
\begin{equation}
  \label{eq:comb_not}
  \left \langle |U_{j,i}U_{j,k}U_{l,k}U_{l,m}U_{n,m}U_{n,i}|^2
  \right\rangle = W^{j,l,n}_{i,k,m}.
\end{equation}
To cut down the number of the coefficients $W$ that need to be
calculated, we list some of the properties arising from the symmetries
of (\ref{eq:avU}).
\begin{enumerate}
\item Application of a permutation to both ${\bf a}$ and ${\bf b}$
  leaves $W$ invariant.  Due to the special form of ${\bf a}$ and
  ${\bf b}$, the only eligible permutations are shifts:
  \begin{equation}
    \label{eq:W_shift_inv}
    W^{j,l,n}_{i,k,m} = W^{n,j,l}_{m,i,k} = W^{l,n,j}_{k,m,i}
  \end{equation}
\item We can switch around ${\bf a}$ and ${\bf b}$.  After that we
  should apply a right shift to bring them back to the special form:
  ${\bf a'}=(i,i,k,k,m,m)$ and ${\bf b'}=(n,j,j,l,l,n)$.  Thus,
  \begin{equation}
    \label{eq:W_swap_inv}
    W^{j,l,n}_{i,k,m} = W_{n,j,l}^{i,k,m}.
  \end{equation}
  Note that the lower indices on the right-hand side are shifted from
  their original order.
\item The coefficients $W$ do not depend on the individual values of
  $j$, $l$ and $n$, but only on whether they are pairwise distinct.
  That is, if
  \begin{equation}
    \label{eq:W_distinct}
    j=l \Leftrightarrow j'=l', \qquad
    l=n \Leftrightarrow l'=n' \quad \mbox{and} \quad
    n=j \Leftrightarrow n'=j',
  \end{equation}
  then 
  \begin{equation}
    \label{eq:W_distinct_eq}
    W^{j,l,n}_{i,k,m} = W^{j',l',n'}_{i,k,m}.
  \end{equation}
  The same applies to $i$, $k$ and $m$.
\end{enumerate}
Using these properties we can list all possible values of $W$.  If
$\{j,l,n\}$ are pairwise distinct, $W^{j,l,n}_{i,k,m} =
W^{1,2,3}_{i,k,m}$ and there are $N(N-1)(N-2)$ ways to select such
$j,l,n$.  If $l=n$ but $j$ is different, $W^{j,l,n}_{i,k,m} =
W^{1,2,2}_{i,k,m}$ and there are $N(N-1)$ way to choose $j$ and $n$;
the cases $j=l\neq n$ and $j=n\neq l$ are analogous.  Finally, when
$j=l=n$, $W^{j,l,n}_{i,k,m} = W^{1,1,1}_{i,k,m}$ and there are $N$
ways to choose the value of $j$.  The dependence on indices
$\{i,k,m\}$ can be treated in the same way.  Table \ref{tab:dubyas}
lists all contributing terms together with the number of times they
occur in expansion (\ref{eq:estim_average}).  The values of $W$
were computed using a program which utilizes formula (\ref{eq:avU})
and the data for $V_N$ from \cite{Sam80}.

\renewcommand{\arraystretch}{1.3}
\begin{table}[h]
  \centering
  \begin{tabular}{|c|c|c|c|}
    \hline \hline
    Contributing terms & Notation & Value & Number of occurrences\\
    \hline \hline
    $W^{1,2,3}_{1,2,3}$ & $w_{33}$ & 
    $\frac{N^5+6N^4+5N^3-20N^2-4N+32}{D(N-2)(N^2-1)(N-1)N}$
    & $N^2(N-1)^2(N-2)^2$ \\[0.2em]
    \hline
    $W^{1,2,2}_{1,2,3}$, $W^{2,1,2}_{1,2,3}$, $W^{2,2,1}_{1,2,3}$
    & $w_{23}$ & 
    $\frac{2(N^2+5N+8)}{D(N-1)N}$ 
    & $N^2(N-1)^2(N-2)$\\[0.2em]
    \hline
    $W^{1,2,2}_{1,2,2}$, $W^{2,1,2}_{2,1,2}$, $W^{2,2,1}_{2,2,1}$ &
    $w_{22a}$ & $\frac{6(N^2+3N+6)}{D(N-1)N}$ & $N^2(N-1)^2$ \\[0.2em]
    \hline
    $W^{1,2,2}_{2,1,2}$, $W^{2,1,2}_{2,2,1}$, $W^{2,2,1}_{1,2,2}$ &
    $w_{22b}$ & $\frac{6(N^2+3N+6)}{D(N-1)N}$ & $N^2(N-1)^2$ \\[0.2em]
    \hline
    $W^{1,2,2}_{2,2,1}$, $W^{2,1,2}_{1,2,2}$, $W^{2,2,1}_{2,1,2}$
    & $w_{22c}$ & 
    $\frac{8(N+2)(N+1)}{D(N-1)N}$ 
    & $N^2(N-1)^2$\\[0.2em]
    \hline
    $W^{1,1,1}_{1,2,3}$ & $w_{13}$ & 
    $\frac{8}{D}$
    & $N^2(N-1)(N-2)$\\[0.2em]
    \hline
    $W^{1,1,1}_{1,2,2}$, $W^{1,1,1}_{2,1,2}$, $W^{1,1,1}_{2,2,1}$ & 
    $w_{12}$ & 
    $\frac{48}{D}$
    & $N^2(N-1)$\\[0.2em]
    \hline
    $W^{1,1,1}_{1,1,1}$ &
    $w_{11}$ & 
    $\frac{720}{D}$
    & $N^2$ \\[0.2em]
    \hline \hline 
  \end{tabular}
  \caption{Factors contributing to
    $\left\langle\Tr\A^3\right\rangle$.  We use notation
    $D=(N+5)(N+4)(N+3)(N+2)(N+1)N$.  The remaining terms can be obtained
    using $W^{j,l,n}_{i,k,m} = W_{n,j,l}^{i,k,m}$.  Whenever the first
    column lists three terms, the last column gives the number of
    occurrences of {\em each} of those terms.}
  \label{tab:dubyas}
\end{table}

Thus we get
\begin{eqnarray*}
  \label{eq:trA_av}
  \fl\left \langle\Tr\A^3\right\rangle_N
  = N^2(N-1)^2(N-2)^2 w_{33} 
  + 2\times3N^2(N-1)^2(N-2) w_{23}\\
  + 3N^2(N-1)^2 w_{22a} + 3N^2(N-1)^2 w_{22b} + 3 N^2(N-1)^2 w_{22c}\\
  + 2\times N^2(N-1)(N-2) w_{13}
  + 2\times3N^2(N-1) w_{12}
  + N^2 w_{11},
\end{eqnarray*}
where the factors of $2$ arise due to $W^{j,l,n}_{i,k,m} =
W^{i,k,m}_{n,j,l}$.   The result is
\begin{eqnarray}
  \label{eq:trA_res}
  \left\langle\Tr\A^3\right\rangle_N
  = \frac{N^6 + 16N^5 + 105N^4 + 370N^3 + 716N^2 + 348N - 136}
  {(N+1)^2(N+5)(N+4)(N+3)(N+2)}\\
  = 1 + 5N^{-2} + O(N^{-3}),
\end{eqnarray}
which, together with inequality $\lambda_2(\A)^3 \leq \Tr\A^3 - 1$,
finishes the proof.

\section{Discussion}
\label{sec:discussion}

We presented an alternative and complete proof of the main result of
\cite{DK04}.  However the main importance of the results of \cite{DK04}
lies in the method of their derivation and in establishing a foothold
in understand the combinatorics of the form factor expansion.  Below
we formulate two conjectures which can be studied using the
methods of \cite{DK04}.

The first conjecture is rephrasing an observation made in \cite{B01}
regarding the asymptotics of the eigenvalues of doubly stochastic matrices
$\M$.
\begin{conj}
  \label{conj:1}
  For any $j\geq2$ and $k$
  $\left\langle|\lambda_j(\M)|^k\right\rangle=O(N^{-k/2})$, where the
  averaging is performed with respect to the measure induced by
  (\ref{eq:trans_prob}).
\end{conj}

The second conjecture is a strengthening of our result which would
follow if Conjecture \ref{conj:1} were proved.  It concerns the speed
of convergence of the diagonal approximation
\begin{conj}
  \label{conj:2}
  For any $\tau\in(0,\infty)$ 
  \begin{equation}
    \label{eq:conj2}
    \left\langle\Tr\A^{\tau N}\right\rangle - 1 \to 0 
  \end{equation}
  faster than any power of $N$.
\end{conj}

\bibliography{all,berkolaiko}
\bibliographystyle{iopart-num}

\end{document}